\documentclass[prl,twocolumn,showpacs,preprintnumbers,amsmath,amssymb,times]{revtex4}

\usepackage{graphicx}% Include figure files
\usepackage{dcolumn}% Align table columns on decimal point
\usepackage{bm}% bold math
\usepackage{psfrag}
\usepackage{epsfig}
\usepackage{amsmath}
\usepackage{amssymb}
\usepackage{color}
\usepackage{fancyheadings}
\usepackage{mathbbol}

\newcommand{\nn}{\nonumber}
\newcommand{\bra}{\langle}

\newcommand{\ket}{\rangle}

\begin{document}

%\preprint{APS/123-QED}

\title{Strongly Interacting Polaritons in Coupled Arrays of Cavities}

\author{Michael J. Hartmann}
\email{m.hartmann@imperial.ac.uk}
\author{Fernando G.S.L. Brand\~ao}
\author{Martin B. Plenio}
\affiliation{Institute for Mathematical Sciences, Imperial College London,
53 Exhibition Road, SW7 2PG, United Kingdom}
\affiliation{QOLS, The Blackett Laboratory, Imperial College London, Prince Consort Road,
SW7 2BW, United Kingdom}

\date{\today}

\begin{abstract}
The experimental observation of quantum phenomena in strongly correlated many
particle systems is difficult because of the short length- and
timescales involved. Obtaining at the same time detailed control of
individual constituents appears even more challenging and thus to date
inhibits employing such systems as quantum computing devices. Substantial progress
to overcome these problems has been achieved with cold atoms
in optical lattices, where a detailed control of collective properties
is feasible but it is very difficult to address and
hence control or measure individual sites. Here we show, that polaritons, combined atom and
photon excitations, in an array of cavities such as a photonic crystal or coupled
toroidal micro-cavities, can form
a strongly interacting many body system, where individual particles can be
controlled and measured. All individual building
blocks of the proposed setting have already been experimentally realised, thus
demonstrating the potential of this device as a quantum simulator.
With the possibility to create attractive on-site potentials the scheme allows for the
creation of highly entangled states and a phase with particles much more delocalised
than in superfluids.
\end{abstract}

%Classification Scheme.
\maketitle

% ---------------------------------------------------------------------------
%
\section{Introduction}

Physical systems where the interactions between the constituent particles are weak and can be treated
perturbatively are mostly well understood. A precise understanding of strongly correlated systems however
appears to be much harder to obtain because the experimental observation of several important mechanisms
is less feasible. Many processes take place at high energy and therefore on fast timescales while
precise control over individual constituent particles is very hard to achieve.
This unattainability of detailed control lies at the heart of the difficulties to experimentally implement
quantum information processes.

Nonetheless, substantial progress has been achieved recently by considering cold atoms trapped in an
optical lattice \cite{JBC+98}, which can effectively be described by a Bose-Hubbard Hamiltonian.
This idea lead to several important experiments emulating condensed matter phenomena.
The most prominent example is probably the observation of the superfluid to Mott insulator
quantum phase transition \cite{GME+02} in a three dimensional lattice, but also other geometries like a Kagome lattice \cite{SBC+04} or a
Tonks Girardeau gas \cite{PWM+04} are being studied. 
Even theoretical ideas for the implementation of
quantum information processing have been pursued by entangling 
neighbouring atoms via controlled collisions \cite{JBC+99,MGW+03}.

In addition, these systems could act as quantum simulators, where the dynamics of a rather complex
Hamiltonian can be simulated in an experiment by eather
creating the Hamiltonian in question, or
Trotter decomposing the evolution into small steps,
where the dynamics of each step is generated by a much simpler Hamiltonian \cite{JZ05,MBZ06}.

Despite their success, optical lattices are in all their applications limited by
an intrinsic draw back: It is exceedingly difficult to access
individual lattice sites, since their separation is only half of the employed optical wavelength.

Here we propose an alternative way to create an effective Bose-Hubbard model and other Hamiltonians, which does not suffer
from this problem. We consider two realisations of an array of cavities and study the dynamics of polaritons,
combined atom photon excitations, in this arrangement. Since the distance between adjacent cavities is considerably larger than the optical wavelength of the resonant mode, individual cavities can be addressed.

\begin{figure}
\includegraphics[width=8cm]{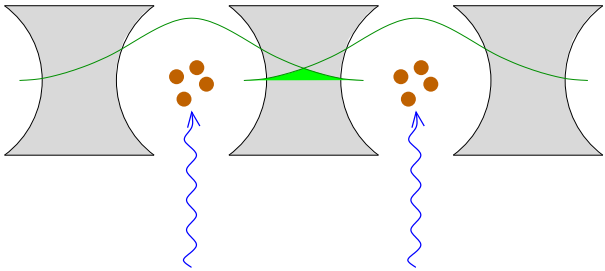}
\caption{\label{crystal} Our model consists of an array of cavities, where photon hopping occurs due to the overlap (shaded green) of the light modes (green lines) of adjacent cavities. Atoms in each cavity (brown), which are driven by external lasers (blue) give
rise to an on site repulsion.}
\end{figure}

Photon hopping occurs between neighboring cavities while
the repulsive force between two polaritons occupying the same site is generated by a large Kerr nonlinearity that occurs if atoms with
a specific level structure usually considered in Electromagnetically Induced Transparency
\cite{HFI90,SI96,ISWD97,WI99,GWG98,Gehri99,Hau99} interact with light.
Each cavity is interacting with an ensemble of these atoms, which are driven by an external laser, see fig. \ref{crystal}.
By varying the intensity of the driving laser, the nonlinearity can be tuned and hence the system can be driven through the superfluid to Mott insulator transition.
In particular, the driving laser may be adjusted for each cavity individually allowing for a much wider
range of tuning possibilities than in an optical lattice, including attractive interactions.

Our system can be described in terms of three species of
polaritons, where under the conditions specified below, the species which is least vulnerable
to decay processes is governed by the effective Bose-Hubbard (BH) Hamiltonian,
\begin{equation} \label{bosehubbard}
H_{\text{eff}} = \kappa \, \sum_{\vec{R}} \, \left(p_{\vec{R}}^{\dagger}\right)^2 \left(p_{\vec{R}}\right)^2 \, + \,
J \sum_{\bra \vec{R}, \vec{R}' \ket} \left(
p_{\vec{R}}^{\dagger} \, p_{\vec{R}'}\, + \, \text{h.c.} \right) \, .
\end{equation}
$p_{\vec{R}}^{\dagger}$ creates a polariton in the cavity at site $\vec{R}$ and the
parameters $\kappa$ and $J$ describe on site repulsion and inter cavity hopping respectively.

The most promising candidates for an experimental realisation are toroidal or spherical micro-cavities,
which are coupled via tapered optical fibres \cite{AKS+03}. These cavities can be produced and positioned with high precision and in large numbers. They have a very large Q-factor ($> 10^8$) for light
that is trapped as whispering gallery modes and efficient coupling to optical fibres \cite{YAV03} as well as coupling to Cs-atoms in close proximity to the cavity via the evanescent field \cite{ADW+06,BPK06} have been demonstrated experimentally very recently.

In the longer term, photonic crystals represent an appealing
alternative as they offer the possibility for the fabrication of
large arrays of cavities in lattices or networks \cite{YXLS99,BHA+05,AAS+03,LSB+04}.
Obviously, our concept is not limited to certain geometries, as it is the case for the optical lattice.
Any arrangement of the cavity array may be considered.

We first derive the Hamiltonian (\ref{bosehubbard}) for the considered structure and then
present a theoretical analysis of the feasible parameter range, which is
backed up by a full numerical demonstration of the superfluid to Mott insulator
transition including experimental imperfections.

\section{Quantised electromagnetic field in a periodic array of cavities}

We consider a periodic array of cavities, which we describe here by a periodic 
dielectric constant,
\begin{equation}
\epsilon (\vec{r}) = \epsilon (\vec{r} + \vec{R}) \, ; \quad \vec{R} = \vec{n} \, R \, ,
\end{equation}
for all tupels of integers $\vec{n} = (n_x, n_y, n_z)$.
We assume the dielectric constant $\epsilon$ to be real, i.e. we neglect absorption processes,
and limit our considerations to linear, isotropic dielectric media.
The electromagnetic field may be represented by a vector potential $\vec{A}$
and a scalar potential $\Phi$ which obey the gauge conditions, $\Phi = 0$ and
$\nabla \cdot (\epsilon(\vec{r}) \, \vec{A}) = 0$ \cite{GL91}.
$\vec{A}$ can be expanded in Wannier functions,
$\vec{w}_{\vec{R}}$, each localised at one single cavity at location
$\vec{R}$. We describe this single cavity by the dielectric function $\epsilon_{\vec{R}}(\vec{r})$ such
that the Wannier functions satisfy the eigenvalue equation,
\begin{equation} \label{Wanniereval}
\frac{\epsilon_{\vec{R}}(\vec{r}) \, \omega_C^2}{c^2} \vec{w}_{\vec{R}} \, -
\, \nabla \times \left( \nabla \times \vec{w}_{\vec{R}} \right) = 0 \, ,
\end{equation}
where the eigenvalue $\omega_C^2$ is the square the resonance frequency of the cavity which is independent of $\vec{R}$ due to the periodicity. 
We assume that the Wannier functions decay strongly enough outside the cavity such that only
Wannier modes of nearest neighbour cavities have nonvanishing overlap.

In terms of the creation and annihilation operators of the Wannier modes, $a_{\vec{R}}^{\dagger}$ and $a_{\vec{R}}$,
the Hamiltonian of the field can be written,
\begin{equation} \label{arrayham2}
\mathcal{H} = \omega_C \sum_{\vec{R}}
\left(  a_{\vec{R}}^{\dagger} a_{\vec{R}} + \frac{1}{2} \right) +
2 \omega_C \alpha \sum_{\bra \vec{R}, \vec{R}' \ket}
\left( a_{\vec{R}}^{\dagger} a_{\vec{R}'} + \text{h.c.} \right) \, .
\end{equation}
Here $\sum_{\bra \vec{R}, \vec{R}' \ket}$ is the sum of all pairs of cavities which are nearest neighbours
of each other.
Since $\alpha \ll 1$, we neglected rotating terms which contain products of two
creation or two annihilation operators of Wannier modes in deriving (\ref{arrayham2}). $\alpha$ is given by \cite{YXLS99,BTO00}, 
\begin{equation}
\alpha = \int d^3r \, \left(\epsilon_{\vec{R}}(\vec{r}) \, - \, \epsilon(\vec{r}) \right) \,
\vec{w}_{\vec{R}}^{\star} \vec{w}_{\vec{R}'} \, ; \quad 
|\vec{R} - \vec{R}'| = R \, ,
\end{equation}
and can be obtained numerically for specific models \cite{MV97}.

The model (\ref{arrayham2}) provides an excellent approximation to many relevant implementations such
as coupled photonic crystal micro-cavities or fibre coupled toroidal micro-cavities.
Furthermore, it allows for the observation of state transfer \cite{HRP06} and entanglement
dynamics and propagation for Gaussian states \cite{PHE04,EPBH04}
as the Hamiltonian is harmonic.
In the next section we present a possible realisation of the repulsive term in the Hamiltonian (\ref{bosehubbard}).

\section{The effective on site repulsion} 

\begin{figure}
\psfrag{g13}{$g_{13}$}
\psfrag{g24}{$g_{24}$}
\psfrag{o}{$\Omega_L$}
\psfrag{d}{$\Delta$}
\psfrag{d2}{\hspace{-0.1cm}$\delta$}
\psfrag{e}{$\varepsilon$}
\psfrag{w}{\hspace{-0.1cm}$\omega_C$}
\psfrag{1}{$1$}
\psfrag{2}{$2$}
\psfrag{3}{$3$}
\psfrag{4}{$4$}
\includegraphics[width=8cm]{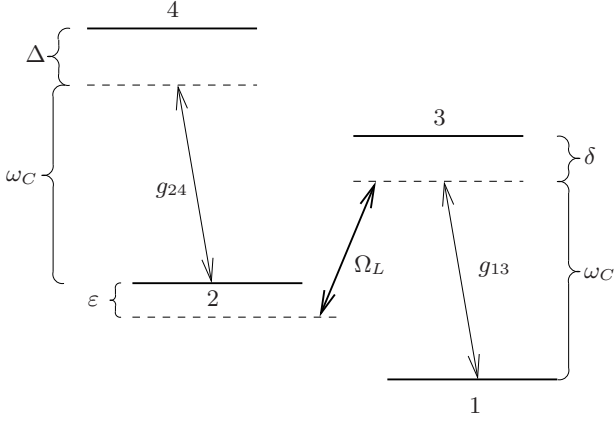}
\caption{\label{level} The level structure and the possible transitions of one atom, $\omega_C$ is the frequency of the cavity mode,
$\Omega_L$ is the Rabi frequency of the driving by the laser, $g_{13}$ and $g_{24}$
are the parameters of the respective dipole couplings and $\delta$, $\Delta$ and $\varepsilon$
are detunings.}
\end{figure}

To generate a repulsion between
polaritons that are located in the same cavity, we fill the cavity with 4 level atoms of a particular level structure that
are driven with an external laser in the same manner as in Electromagnetically Induced Transparency,
see figure \ref{level}:
The transitions between levels 2 and 3 are coupled to the laser field
and the transitions between levels 2-4 and 1-3 couple via dipole moments to the cavity resonance mode.
It has been shown by Imamoglu and co-workers, that this atom cavity system can show a very large nonlinearity \cite{ISWD97}, and a similar nonlinearity has recently been observed experimentally \cite{BBM+05}.

In a rotating frame with respect to
$H_0 = \omega_C \left( a^{\dagger} a + \frac{1}{2} \right) + \sum_{j=1}^N \left( \omega_C \sigma_{22}^j + \omega_C \sigma_{33}^j + 2 \omega_C \sigma_{44}^j \right)$,
the Hamiltonian of the atoms in the cavity reads,
\begin{eqnarray} \label{H_manyatom}
H_I & = &
\sum_{j=1}^N \left(\varepsilon \sigma_{22}^j + \delta \sigma_{33}^j + (\Delta + \varepsilon)
\sigma_{44}^j \right) \\
& + & \, \sum_{j=1}^N \left( \Omega_L \, \sigma_{23}^j \, + \,
g_{13} \, \sigma_{13}^j \, a^{\dagger} \, + \,
g_{24} \, \sigma_{24}^j \, a^{\dagger} \, + \, \text{h.c.} \right) \nn \, ,
\end{eqnarray}
where $\sigma_{kl}^j = | k_j \ket \bra l_j |$ projects level $l$ of atom $j$ to level $k$ of the same atom,
$\omega_C$ is the frequency of the cavity mode,
$\Omega_L$ is the Rabi frequency of the driving by the laser and $g_{13}$ and $g_{24}$
are the parameters of the dipole coupling of the cavity mode to the respective atomic
transitions which are all assumed to be real.

Since we are only interested in the situation, where there is on average one excitation in each
cavity and since the probability to have two excitations in one cavity is suppressed by the Kerr
nonlinearity, it is appropriate to only consider states with at most two
excitations per cavity. All atoms interact in the same way with the cavity mode and hence the only
relevant states are Dicke type dressed states.
Neglecting two photon detuning and coupling to level 4, the Hamiltonian (\ref{H_manyatom}) can
be written as a model of three species of polaritons in this subspace.

\subsection{Polaritons}

In the case where $g_{24} = 0$ and $\varepsilon = 0$ , level 4 of the atoms decouples from the rest
of the two excitation manifold \cite{RPT02,WI99}.
If we furthermore assume that the number of atoms is large, $N \gg 1$, the Hamiltonian (\ref{H_manyatom}) truncated to the subspace of at most two excitations can be diagonalised.
Let us therefore define the following creation (and annihilation) operators:
\begin{eqnarray} \label{polariton_operators}
p_0^{\dagger} & = & \frac{1}{B} \, \left(g S_{12}^{\dagger} - \Omega_L a^{\dagger} \right) \nn \\
p_+^{\dagger} & = & \sqrt{\frac{2}{A (A + \delta)}} \, \left(\Omega_L S_{12}^{\dagger} + g a^{\dagger} +
\frac{A + \delta}{2} S_{13}^{\dagger} \right) \nn \\
p_-^{\dagger} & = & \sqrt{\frac{2}{A (A - \delta)}} \, \left(\Omega_L S_{12}^{\dagger} + g a^{\dagger} -
\frac{A - \delta}{2} S_{13}^{\dagger} \right) \, ,
\end{eqnarray}
where $g = \sqrt{N} g_{13}$, $B = \sqrt{g^2 + \Omega_L^2}$, $A = \sqrt{4 B^2 + \delta^2}$,
$S_{12}^{\dagger} = \frac{1}{\sqrt{N}} \sum_{j=1}^N \sigma_{21}^j$
and $S_{13}^{\dagger} = \frac{1}{\sqrt{N}} \sum_{j=1}^N \sigma_{31}^j$.

The operators $p_0^{\dagger}$, $p_+^{\dagger}$ and $p_-^{\dagger}$ describe polaritons,
quasi particles formed by combinations of atom and photon excitations.
By looking at their matrix representation in the subspace of at most two excitations,
one can see that in the limit of large atom numbers, $N \gg 1$, they satisfy bosonic commutation relations,
\begin{equation} \label{polariton_comm}
\left[ p_j, p_l \right] = \, 0 \quad ; \quad
\left[ p_j, p_l^{\dagger} \right] = \delta_{jl} \quad ; \quad j,l = 0,+,- \, . 
\end{equation}
$p_0^{\dagger}$, $p_+^{\dagger}$ and $p_-^{\dagger}$ thus describe independent bosonic
particles. In terms of these polaritons, the Hamiltonian (\ref{H_manyatom}) for $g_{24} = 0$ and
$\varepsilon = 0$ reads,
\begin{equation} \label{H0polariton}
\left[H_I\right]_{g_{24} = 0, \varepsilon = 0} = 
\mu_0 \, p_0^{\dagger} p_0 + \mu_+ \, p_+^{\dagger} p_+ + \mu_- \, p_-^{\dagger} p_- \, ,
\end{equation}
where the frequencies are given by $\mu_0 = 0$, $\mu_+ = (\delta - A)/2$ and
$\mu_- = (\delta + A)/2$. We will now focus on the dark state polaritons, $p_0^{\dagger}$,
which can be described by the Hamiltonian (\ref{bosehubbard}) as we shall see.

\subsection{Perturbations\label{perturbations}} 

To write the full Hamiltonian $H_I$, (\ref{H_manyatom}), in the polariton basis,
we express the operators $\sum_{j=1}^N \sigma_{22}^j$ and $a^{\dagger} \,
\sum_{j=1}^N \sigma_{24}^j$
in terms of $p_0^{\dagger}$, $p_+^{\dagger}$ and $p_-^{\dagger}$. 

In the subspace of at most two excitations, the coupling
of the polaritons to the level 4 of the atoms via the dipole moment $g_{24}$ reads,
\begin{equation} \label{coupletolevel4}
g_{24} \, \left( \sum_{j=1}^N \sigma_{42}^j \, a \, + \text{h.c.} \right) \approx
- g_{24} \, \frac{g \Omega_L}{B^2} \, \left( S_{14}^{\dagger} \, p_0^2 \, + \text{h.c.} \right) \, ,
\end{equation}
where $S_{14}^{\dagger} = \frac{1}{\sqrt{N}} \sum_{j=1}^N \sigma_{41}^j$. 
In deriving (\ref{coupletolevel4}), we made use of the rotating wave approximation: In a
frame rotating with respect to (\ref{H0polariton}), the polaritonic creation operators rotate with the frequencies $\mu_0$, $\mu_+$ and $\mu_-$.
Furthermore, the operator $S_{14}^{\dagger}$ rotates at the frequency $2 \mu_0$, (c.f. (\ref{H_manyatom})). Hence, provided that 
\begin{equation} \label{rotatingwappr}
|g_{24}| \, , \, |\varepsilon| \, , \, |\Delta| \, \ll \, |\mu_+| \, , \, |\mu_-|
\end{equation}
all terms that rotate at
frequencies $2 \mu_0 - (\mu_+ \pm \mu_-)$ or $\mu_0 \pm \mu_+$ or $\mu_0 \pm \mu_-$
can be neglected, which eliminates all interactions that would couple $p_0^{\dagger}$ and $S_{14}^{\dagger}$
to the remaining polariton species.

For $g_{24} \ll |\Delta|$, the coupling to level 4 can be treated in a perturbative way.
This results in an energy shift of $2 \, \kappa$ with
\begin{equation}
\kappa = - \frac{g_{24}^2}{\Delta} \,
\frac{N g_{13}^2 \, \Omega_L^2}{\left(N g_{13}^2 \, + \, \Omega_L^2 \right)^2}
\end{equation}
and in an occupation probability of the state of one $S_{14}^{\dagger}$ excitation of
$- 2 \kappa / \Delta$, 
which will determine an effective decay rate for the polariton $p_0^{\dagger}$ via
spontaneous emission from level 4 (see below).
Note that $\kappa > 0$ for $\Delta < 0$ and vice versa.
In a similar way, the two photon detuning $\varepsilon$ leads to and energy shift of
$\varepsilon \, g^2 \, B^{-2}$ for the polariton $p_0^{\dagger}$, which plays the role of a chemical
potential in the effective Hamiltonian.

Hence, provided (\ref{rotatingwappr}) holds, the Hamiltonian for the dark state polariton $p_0^{\dagger}$
can be written as
\begin{equation} \label{Heffect}
H_{\text{eff}} = \kappa \, \left(p_0^{\dagger}\right)^2 \left(p_0\right)^2 \, + \, \varepsilon \, \frac{g^2}{B^2} \, p_0^{\dagger} p_0 \, ,
\end{equation}
in the rotating frame. Next we turn to estimate the effects of loss and decoherence.

\subsection{Spontaneous emission and cavity decay}

Since level 2 of the atoms is metastable and hence its decay rate negligible on the relevant time scales,
the dark state polaritons $p_0^{\dagger}$ suffer loss and decoherence via two processes only:
The photons are subject to cavity decay at a rate $\Gamma_C$ and spontaneous emission from the atomic level 4 at a rate $\Gamma_4$ induces an effective decay via its small but nonvanishing occupation.

Taking into account the definition (\ref{polariton_operators}) of $p_0^{\dagger}$ and the
effective decay of $p_0^{\dagger}$ induced by the occupation of level 4,
we obtain the following decay rate $\Gamma$ for the polaritons $p_0^{\dagger}$:
\begin{equation}
\Gamma = \frac{2 \, \Omega_L^2}{N g_{13}^2 \, + \, \Omega_L^2} \, \left(
\frac{\Gamma_C}{2} \, + \, \frac{g_{24}^2 N g_{13}^2}{\Delta^2 \left(N g_{13}^2 \, + \, \Omega_L^2 \right)} \, \Gamma_4 \right) \, .
\end{equation}
Due to $g_{24} \ll |\Delta|$, spontaneous emission from level 4 affects the polaritons much less than
cavity decay.

For successfully observing the dynamics and phases of the effective Hamiltonian
(\ref{Heffect}), the repulsion term $\kappa$
needs to be much larger than the damping $\Gamma$.
Choosing $\Omega_L \ll \sqrt{N} g_{13}$, one can achieve a ratio of
$\kappa / \Gamma \sim 0.4$ for photonic band gap cavities \cite{SKV+05},
while for toroidal micro-cavities the much larger value of $\kappa / \Gamma \sim 1.1 \times 10^{3}$
is possible, which makes them an ideal candidate for an experimental implementation \cite{SKV+05,YAV03,ADW+06}.
In the next section we turn to combine the on site repulsion for polaritons with the
photon hopping between neighbouring cavities. 

\section{The complete picture}

Now we look at an array of cavities which forms a periodic dielectric medium, see fig. \ref{crystal}.
The first term of the Hamiltonian
(\ref{arrayham2}) has already been incorporated in the polariton analysis for one individual cavity.
The second term transforms into the polariton picture via (\ref{polariton_operators}).
To distinguish between the dark state polaritons in different
cavities, we introduce the notation $p_{\vec{R}}^{\dagger}$ to label the polariton $p_0^{\dagger}$
in the cavity at position $\vec{R}$. We get,
\begin{eqnarray} 
a_{\vec{R}}^{\dagger} a_{\vec{R}'}
& \approx & \frac{\Omega_L^2}{B^2} \, p_{\vec{R}}^{\dagger} \, p_{\vec{R}'} \\
& + & \text{"terms for other polariton species"} \nn \, ,
\end{eqnarray}
where we have applied a rotating wave approximation.
Contributions of different polaritons decouple due to the separation of their
frequencies $\mu_0$, $\mu_+$ and $\mu_-$.
As a consequence the Hamiltonian for the polaritons $p_{\vec{R}}^{\dagger}$ takes on the form
(\ref{bosehubbard}), with 
\begin{equation}
J = \frac{2 \omega_C \Omega_L^2}{N g_{13}^2 + \Omega_L^2} \alpha \, ,
\end{equation}
where we have assumed a negligible two photon detuning, $\varepsilon \approx 0$.

\section{Numerical Analysis}

To provide evidence for the accuracy of our approach, we present a numerical simulation of the
dynamics of one dark state polariton in an array of three toroidal micro-cavities with periodic boundary
conditions. Parameters from \cite{SKV+05},
$g_{13} = 2.5 \times 10^{9} \text{s}^{-1}$,
$g_{24} = 2.5 \times 10^{9} \text{s}^{-1}$,
$\Gamma_4 = 1.6 \times 10^{7} \text{s}^{-1}$,
$\Gamma_C = 0.4 \times 10^{5} \text{s}^{-1}$, and assuming
$\delta = 1.0 \times 10^{12} \text{s}^{-1}$,
$\Delta = - 1.0 \times 10^{11} \text{s}^{-1}$,
$N = 10000$, $\varepsilon = 0$, $2 \omega_C \alpha = 0.4 \times 10^{8} \text{s}^{-1}$ and
$\Omega_L = 2.5 \times 10^{13} \text{s}^{-1}$ yields values of
$\kappa = 1.6 \times 10^{7} \text{s}^{-1}$ and
$J = 2.0 \times 10^{7} \text{s}^{-1}$.

\begin{figure}
\psfrag{t}{\hspace{-0.4cm}\raisebox{-0.3cm}{\tiny $t$ in $10^{-8}$ s}}
\psfrag{data1}{\hspace{0.01cm} \tiny $n_1$}
\psfrag{data2}{\hspace{0.01cm} \tiny $n_2$}
\psfrag{data3}{\hspace{0.01cm} \tiny $\Delta_1$}
\psfrag{data4}{\hspace{0.01cm} \tiny $\Delta_2$}
\psfrag{0}{\raisebox{-0.1cm}{\tiny $0$}}
\psfrag{2}{\raisebox{-0.1cm}{\tiny $2$}}
\psfrag{4}{\raisebox{-0.1cm}{\tiny $4$}}
\psfrag{6}{\raisebox{-0.1cm}{\tiny $6$}}
\psfrag{8}{\raisebox{-0.1cm}{\tiny $8$}}
\psfrag{10}{\raisebox{-0.1cm}{\tiny $10$}}
\psfrag{12}{\raisebox{-0.1cm}{\tiny $12$}}
\psfrag{x 10}{\tiny $10^{-4}$}
\psfrag{-4}{$ $}
\psfrag{0a}{\hspace{-0.2cm}\tiny $0$}
\psfrag{0.2a}{\hspace{-0.2cm}\tiny $0.2$}
\psfrag{0.4a}{\hspace{-0.2cm}\tiny $0.4$}
\psfrag{0.6a}{\hspace{-0.2cm}\tiny $0.6$}
\psfrag{0.8a}{\hspace{-0.2cm}\tiny $0.8$}
\psfrag{1a}{\hspace{-0.2cm}\tiny $1$}
\psfrag{data1b}{\hspace{0.1cm} \tiny $\delta n_1$}
\psfrag{data2b}{\hspace{0.1cm} \tiny $\delta n_2$}
\psfrag{data3b}{\hspace{0.1cm} \tiny $\delta \Delta_1$}
\psfrag{data4b}{\hspace{0.1cm} \tiny $\delta \Delta_2$}
\psfrag{-2b}{\hspace{-0.2cm}\tiny $-2$}
\psfrag{0b}{\hspace{-0.1cm}\tiny $0$}
\psfrag{2b}{\hspace{-0.1cm}\tiny $2$}
\psfrag{4b}{\hspace{-0.1cm}\tiny $4$}
\includegraphics[width=6cm]{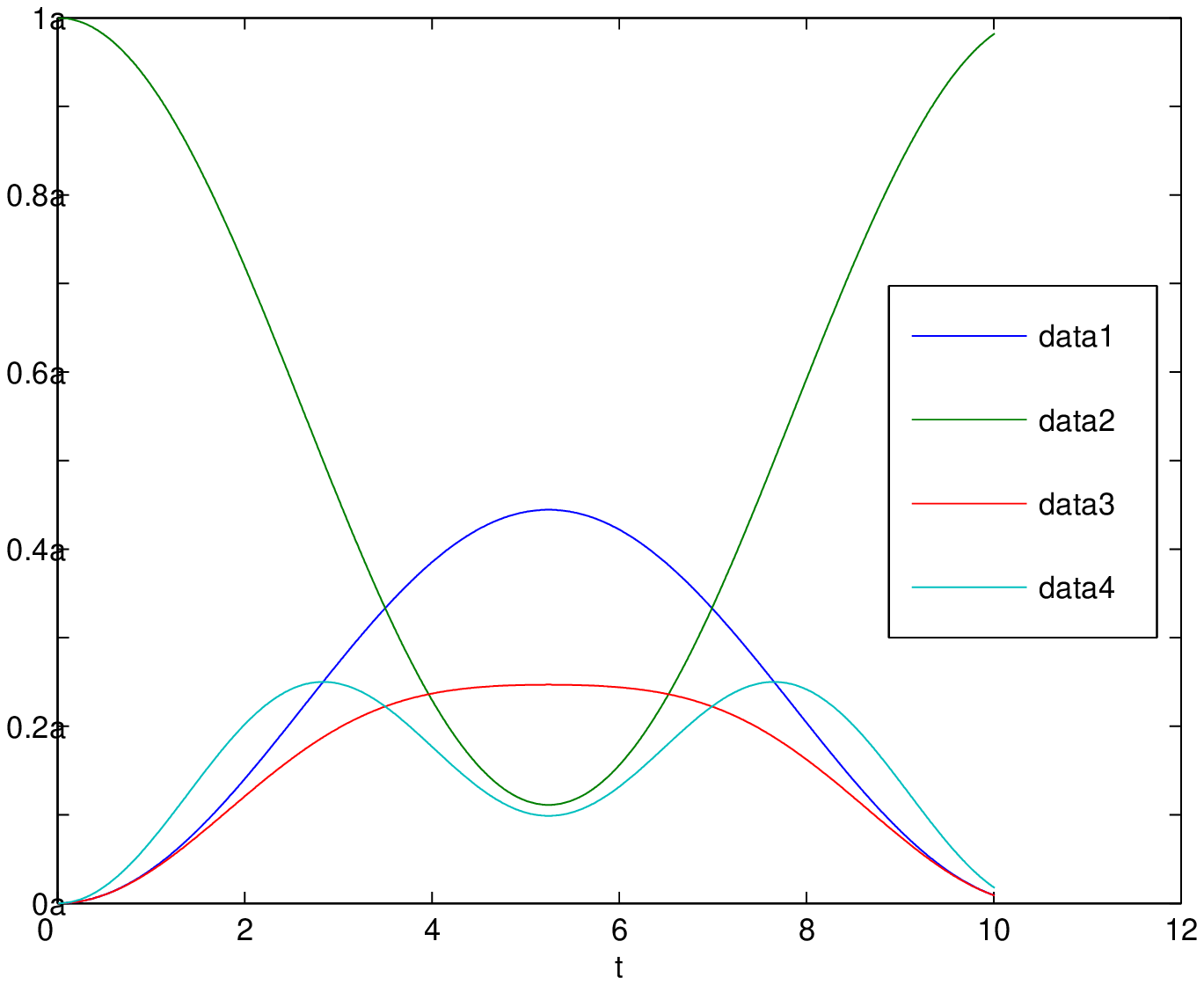}\\[0.2cm]
\includegraphics[width=6cm]{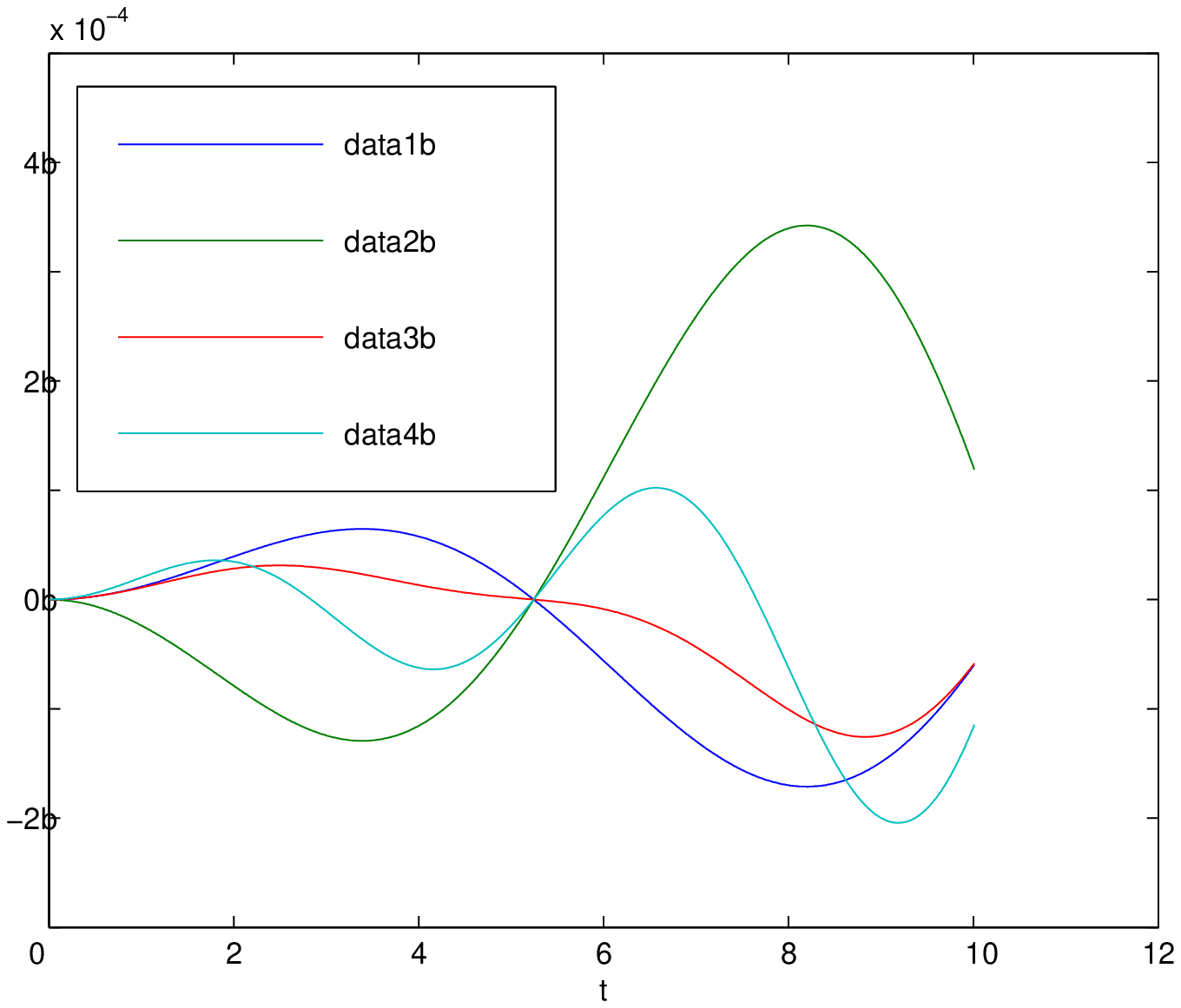}
\caption{\label{dynamics} Upper plot: Cavity polariton numbers and number fluctuations for cavities 1 and 2.
Lower plot: Differences between the cavity polariton number in one cavity
and the excitation number at one site for the pure BH model, $\delta n_l$,
and difference between both models in the number fluctuations, $\delta \Delta_l$ for cavities/sites 1 and 2. Deviations from the pure BH model are less than $10^{-3}$.}
\end{figure}

The polariton is initially located in cavity 2,
and we compare its dynamics to that of a pure BH model.
In doing so we look at the number of polaritons in one cavity/site, 
$n_l = \langle p_l^{\dagger} p_l \rangle$ for site $l$, and the number fluctuations,
$\Delta_l = \langle (p_l^{\dagger} p_l )^2 \rangle - \langle p_l^{\dagger} p_l \rangle^2$.
Figure \ref{dynamics} shows cavity polariton numbers and number fluctuations for cavities 1 and 2, in the upper plot while the lower plot
shows the differences between the cavity polariton number in one cavity
and the excitation number at one site for the pure BH model,
$\delta n_l = \left[n_l\right]_{\text{cavities}} - \left[n_l\right]_{\text{BH}}$, as well
as the difference between both models in the number fluctuations,
$\delta \Delta_l = \left[\Delta_l\right]_{\text{cavities}} - \left[\Delta_l\right]_{\text{BH}}$.
All our simulation results show a single trajectory of a quantum jump simulation \cite{PK98}. Within the simulated time range the probability of a decay event remained less than $2 \times 10^{-3}$.

A key question is of course, whether the polaritons undergo the Mott insulator to superfluid
phase transition in the same way as a pure BH model. We give evidence for that in the next section.

\subsection{The phase transition}

An important feature of the Bose-Hubbard Hamiltonian (\ref{bosehubbard}) are two
phases of the matter it describes: One is a superfluid phase, where the hopping term dominates over the
on site repulsion, $J \gg \kappa$, and the polaritons $p_0^{\dagger}(\vec{R})$ move almost freely between the cavities. As a consequence one can observe large fluctuations in the number of polaritons in each
individual cavity.
The other phase is the Mott insulator, in which the on site repulsion is much stronger than the hopping
term $J \ll \kappa$ and double occupancies of a cavity are strongly suppressed.
Provided the system contains on average one polariton per cavity, this implies that in the Mott phase the number of polaritons is exactly one for each cavity and number fluctuations are very small.

\begin{figure}
\psfrag{data1}{\hspace{0.12cm} \small $\kappa$}
\psfrag{data2}{\hspace{0.12cm} \small $J$}
\psfrag{t}{\hspace{-0.4cm}\raisebox{-0.3cm}{\tiny $t$ in $10^{-6}$ s}}
\psfrag{h}{\hspace{-0.2cm}\raisebox{0.4cm}{\tiny MHz}}
\psfrag{0}{\raisebox{-0.06cm}{\tiny $0$}}
\psfrag{200}{\raisebox{-0.06cm}{\tiny $0.2$}}
\psfrag{400}{\raisebox{-0.06cm}{\tiny $0.4$}}
\psfrag{600}{\raisebox{-0.06cm}{\tiny $0.6$}}
\psfrag{800}{\raisebox{-0.04cm}{\tiny $0.8$}}
\psfrag{1000}{\raisebox{-0.04cm}{\tiny $1.0$}}
\psfrag{-0.08a}{\hspace{-0.31cm}\tiny $-0.08$}
\psfrag{-0.06a}{\hspace{-0.31cm}\tiny $-0.06$}
\psfrag{-0.04a}{\hspace{-0.31cm}\tiny $-0.04$}
\psfrag{-0.02a}{\hspace{-0.31cm}\tiny $-0.02$}
\psfrag{0a}{\hspace{-0.12cm}\tiny $0$}
\psfrag{0.02a}{\hspace{-0.22cm}\tiny $0.02$}
\psfrag{0.04a}{\hspace{-0.22cm}\tiny $0.04$}
\psfrag{0.06a}{\hspace{-0.22cm}\tiny $0.06$}
\psfrag{0.1a}{\hspace{-0.15cm}\tiny $0.1$}
\psfrag{0.2a}{\hspace{-0.15cm}\tiny $0.2$}
\psfrag{0.3a}{\hspace{-0.15cm}\tiny $0.3$}
\psfrag{0.4a}{\hspace{-0.15cm}\tiny $0.4$}
\psfrag{0.5a}{\hspace{-0.15cm}\tiny $0.5$}
\psfrag{10.0}{\hspace{-0.16cm}\tiny $10^2$}
\psfrag{10.-1}{\hspace{-0.18cm}\tiny $10^1$}
\psfrag{10.-2}{\hspace{-0.18cm}\tiny $10^0$}
\psfrag{10.-3}{\hspace{-0.3cm}\tiny $10^{-1}$}
\includegraphics[width=6cm]{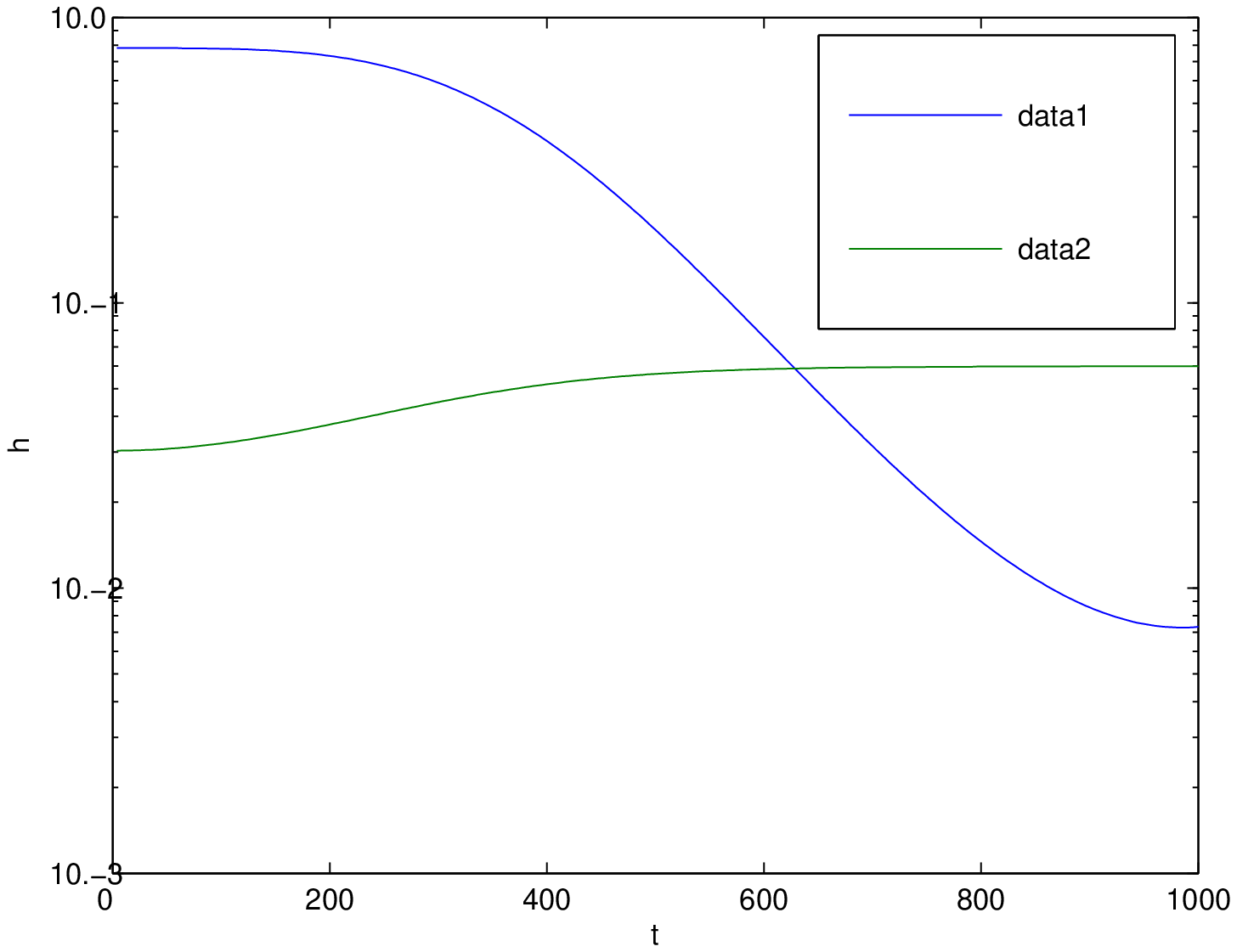}\\[0.2cm]
\includegraphics[width=6cm]{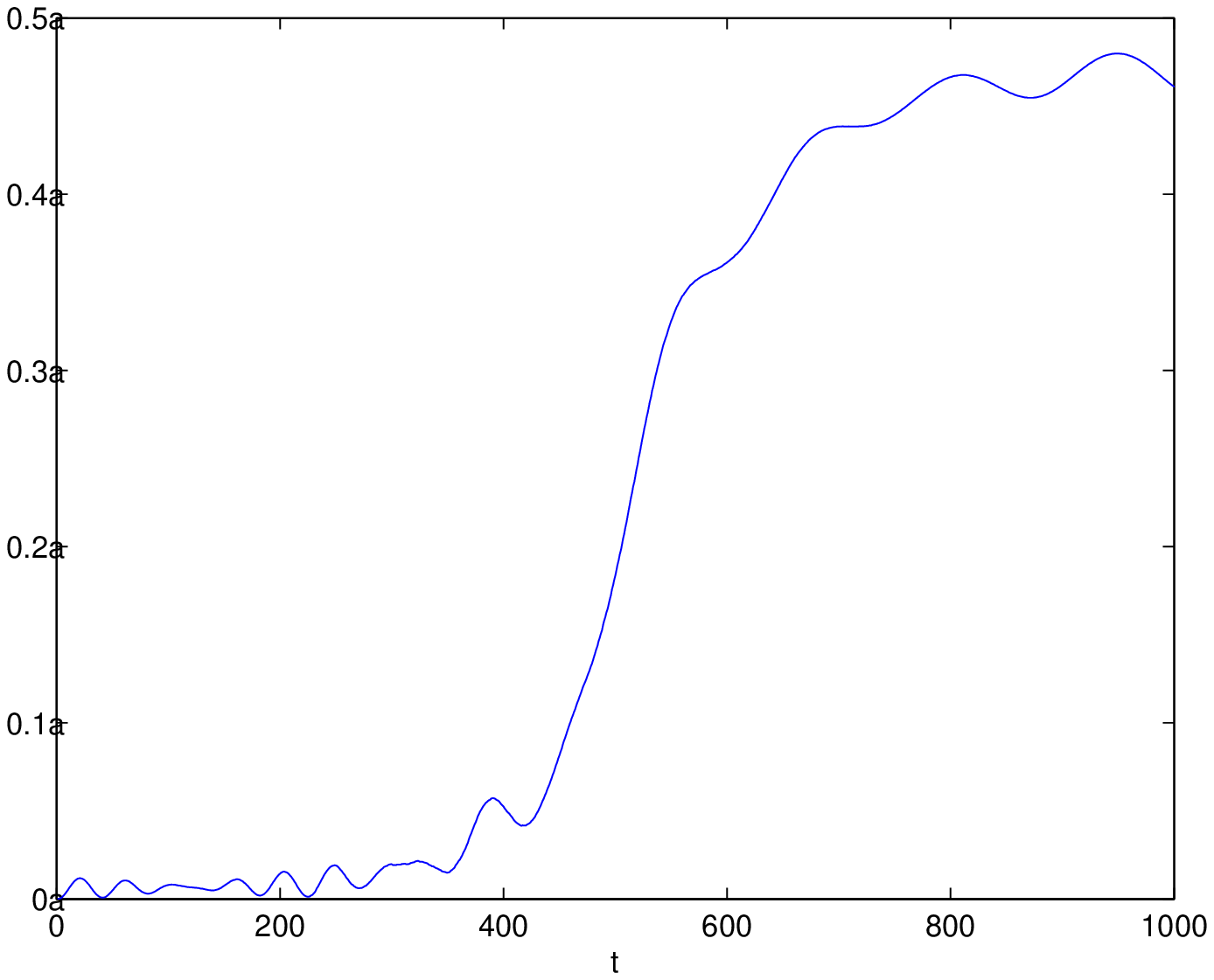}\\[0.2cm]
\includegraphics[width=6cm]{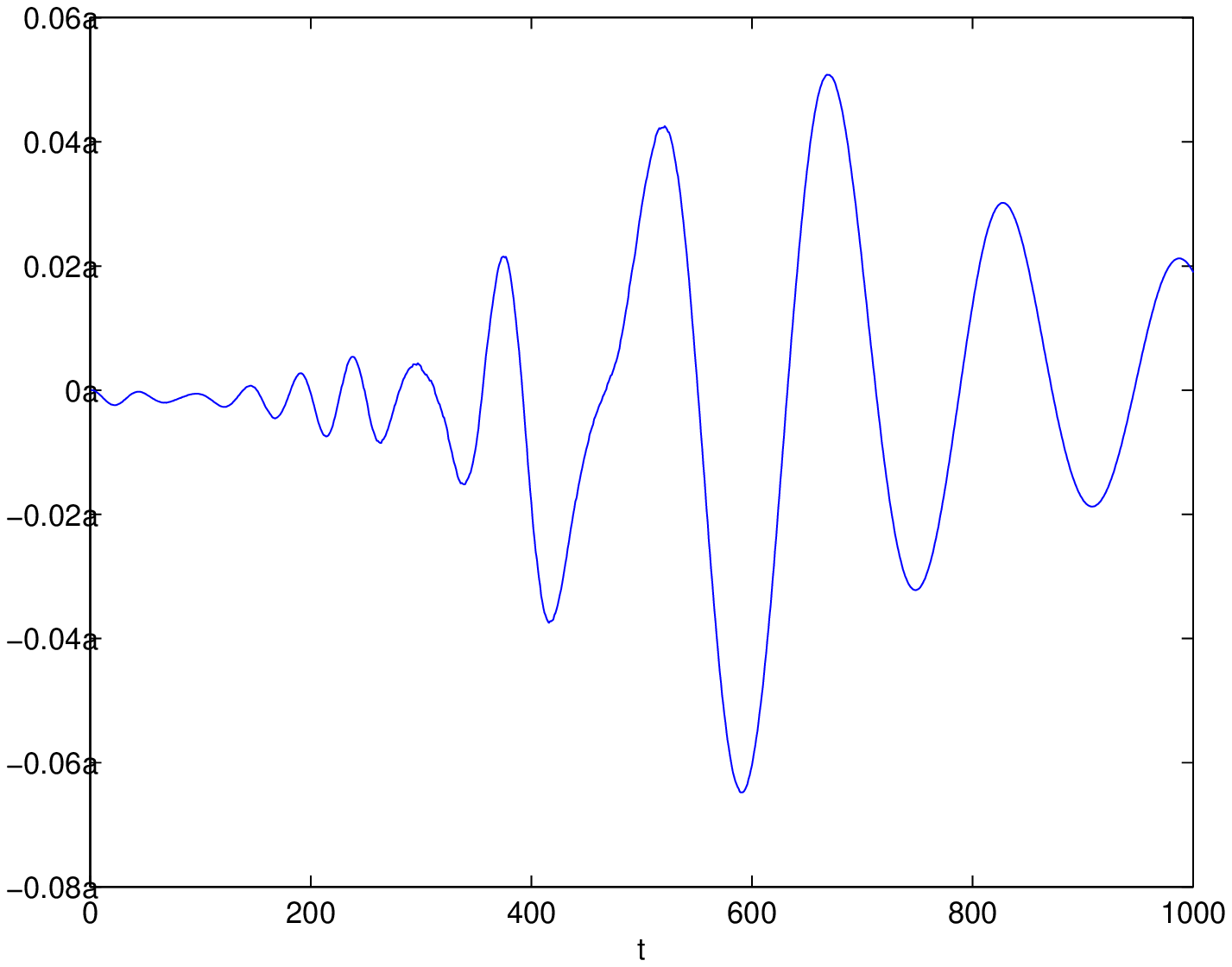}
\caption{\label{vardiff4} Log-plot of $\kappa$ (blue) and $J$ (green) as a function of time, upper plot,
number fluctuations of cavity polaritons in cavity 1, $\Delta_1$,
as a function of time, middle plot and difference between number fluctuations in cavity polaritons and number fluctuations in a pure BH model, $\delta \Delta_1$, as a function of time, lower plot.
Parameters are the same as in figure \ref{dynamics} except for $N = 1000$,
$\Delta = -0.2 \times 10^{11} \text{s}^{-1}$ and $2 \omega_C \alpha = 0.12 \times 10^{8} \text{s}^{-1}$.
Deviations from the pure BH model are about 5\%.}
\end{figure}

A numerical simulation of this phase transition is shown in figure \ref{vardiff4}.
Again we consider the parameters for coupled toroidal micro-cavities,
but now we assume that the driving laser is tuned from very weak driving to strong driving
throughout the experiment.
We look at four coupled cavities with periodic boundary conditions,
which are initially prepared such that there is one polariton $p_0^{\dagger}$ in each.
Figure \ref{vardiff4} shows how the values of $\kappa$ and $J$ vary in time due to the tuning (upper plot),
the number fluctuations of cavity polaritons in cavity 1, $\Delta_1$ (middle plot),
and the difference between number fluctuations in cavity polaritons and number fluctuations in a pure BH model, $\delta \Delta_1$ (lower plot).
The expectation value for the number of polaritons, $\langle p_0^{\dagger} p_0 \rangle$ would be equal
to one for an exact BH model. This also holds for the two cavities
with deviations less than $3 \times 10^{-3}$. 
The probability for a decay event is about 15\% for the total simulated time range.
Next we mention some applications of the setup when operated with attractive on-site potential.

\subsection{Attractive on-site potential}

As mentioned above, if the detuning $\Delta$ is positive, the on-site potential becomes attractive.
The ground state of the system with $N$ polaritons on $N$ sites and a strong attractive on-site potential
is then a highly entangled $W_N$ state;
$| W_N \ket = \frac{1}{\sqrt{N}} \sum_{j=1}^N | N_j \ket$, where $| N_j \ket$ is the state with all
$N$ polaritons on site $j$ \cite{JY05}. Although the ground state becomes degenerate in the limit
$\kappa \rightarrow - \infty$, one can generate the $W_N$ state with high fidelity since the hopping
terms do not induce transitions between the nearly degenerate low energy states.
The procedure is robust against disorder as long as the hopping $J$ is stronger than
the fluctuations of $\kappa$.
In the $W_N$ state regime, the polaritons enter a physically different phase, where the number fluctuations at one site become $\Delta_i = N - 1$.

\begin{figure}
\psfrag{data1}{\hspace{0.04cm} \small $o_{\text{gs}}$}
\psfrag{data2}{\hspace{0.04cm} \small $o_W$}
\psfrag{data3}{\hspace{0.04cm} \small $\Delta_1$}
\psfrag{t}{\hspace{-0.4cm}\raisebox{-0.3cm}{\tiny $t$ in $10^{-6}$ s}}
\psfrag{0}{\raisebox{-0.06cm}{\tiny $0$}}
\psfrag{20}{\raisebox{-0.06cm}{\tiny $0.2$}}
\psfrag{40}{\raisebox{-0.06cm}{\tiny $0.4$}}
\psfrag{60}{\raisebox{-0.06cm}{\tiny $0.6$}}
\psfrag{80}{\raisebox{-0.06cm}{\tiny $0.8$}}
\psfrag{100}{\raisebox{-0.06cm}{\tiny $1.0$}}
\psfrag{120}{\raisebox{-0.06cm}{\tiny $1.2$}}
\psfrag{140}{\raisebox{-0.06cm}{\tiny $1.4$}}
\psfrag{160}{\raisebox{-0.06cm}{\tiny $1.6$}}
\psfrag{180}{\raisebox{-0.06cm}{\tiny $1.8$}}
\psfrag{200}{\raisebox{-0.06cm}{\tiny $2.0$}}
\psfrag{0a}{\hspace{-0.12cm}\tiny $0$}
\psfrag{0.5a}{\hspace{-0.15cm}\tiny $0.5$}
\psfrag{1a}{\hspace{-0.15cm}\tiny $1$}
\psfrag{1.5a}{\hspace{-0.15cm}\tiny $1.5$}
\psfrag{2a}{\hspace{-0.15cm}\tiny $2$}
\psfrag{2.5a}{\hspace{-0.15cm}\tiny $2.5$}
\psfrag{3a}{\hspace{-0.15cm}\tiny $3$}
\includegraphics[width=6cm]{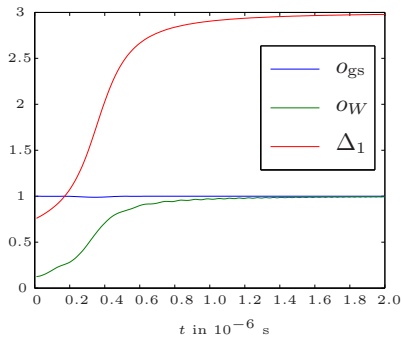}
\caption{\label{entstate} Overlap
with the momentary ground state $o_{\text{gs}} = | \bra \phi (t) | \phi_{\text{gs}} (t) \ket |$, overlap
with the $W_N$ state $o_{W} = | \bra \phi (t) | W_N \ket |$ and the number fluctuations $\Delta_1$
for the effective Hamiltonian (\ref{bosehubbard}) with 4 cavities 
where $J = 10^{7} \text{s}^{-1}$ and $\kappa$ is linearly tuned form
$\kappa_{\text{min}} = -2 \times 10^{5} \text{s}^{-1}$
to $\kappa_{\text{max}} = -4 \times 10^{7} \text{s}^{-1}$.}
\end{figure}

Figure \ref{entstate} shows results of a simulation for the effective Hamiltonian (\ref{bosehubbard})
with 4 cavities where $J = 10^{7} \text{s}^{-1}$ and $\kappa$ is tuned form
$\kappa_{\text{min}} = -2 \times 10^{5} \text{s}^{-1}$
to $\kappa_{\text{max}} = -4 \times 10^{7} \text{s}^{-1}$. The plotted quantities are the overlap
with the momentary ground state $| \bra \phi (t) | \phi_{\text{gs}} (t) \ket |$, the overlap
with the $W_N$ state $| \bra \phi (t) | W_N \ket |$ and the number fluctuations $\Delta_1$.

Switching between positive and negative on-site potential is possible since the detuning $\Delta$ can be varied via the Stark shift induced by an applied electric field.
We now turn to discuss how to distinguish different phases in a measurement.

\section{Measuring the state of the polaritons}

A possible way to detect whether or not there is a polariton present in a certain cavity is to make use
of resonance fluorescence. To that end an additional probe laser is resonantely coupled to the transition
between an atomic level 5, which is well separated from the levels 1 - 4,
and the level 2. The intensity of the probe laser needs to be weak enough such that its
Rabi frequency $\Omega_p$ fulfils $\Omega_p \ll (|\mu_+ - \mu_0|, |\mu_- - \mu_0|)$.
In that case the probe laser couples a polariton $p_0^{\dagger}$ to the extra level 5 while its coupling
to the other polariton species is negligible.
Hence whenever resonant light is detected, at least one polariton $p_0^{\dagger}$ must have been present
in the cavity.

Provided there is on average one polariton per cavity, repeated measurements at several cavities
reveal whether there has always been a polariton in each cavity (Mott phase) or whether there are
every now and then empty cavities indicating polariton number fluctuations (superfluid phase).

Furthermore the fluorescence spectrum contains information
about the energy differences between level 5 and the eigenstates of the polaritonic BH model.
In the limit of negligible Rabi frequency of the probe laser and negligible linewidth of level 5,
the exact spectrum of (\ref{bosehubbard}), which is gapped in the Mott phase and gapless in the superfluid phase, could be observed.
In the real situation, the finite Rabi frequency of the probe laser has to be taken into account and the linewidth of level 5 limits the accuracy of the measurement.

\section{Conclusions}

We have shown, that polaritons, combined atom - photon excitations, can form an effective
quantum many particle system, which is described by a Bose-Hubbard Hamiltonian,
and numerically demonstrated that an observation of the Mott insulator to superfluid phase
transition is indeed feasible for experimentally realisable parameters.  

In contrast to earlier realisations in optical lattices, our scenario has the advantage that
single sites of the BH model can be addressed individually.
This feature opens up various new experimental possibilities: One can for example operate one part
of the system in the Mott insulator regime while the rest is in the superfluid phase and study the
physics that happens at the boundary. This boundary may even be moved around during the experiment.
Local properties of the system may be probed \cite{HMH04,H06}.
Moreover being able to address individual sites is a prerequisite for the implementation of
quantum information processes. Hence we expect our proposal to lead to significant progress in that
direction \cite{GM06,A06}.

In addition, our setup offers the possibility to create a BH type Hamiltonian with an attractive
on site potential, which is readily achieved by choosing the detuning $\Delta$ to be positive,
and may be used to generate highly entangled
states \cite{HBP06}. Even designing many body systems with long range interactions seems feasible since toroidal micro-cavities
are usually coupled to light in optical fibres \cite{YAV03}.
This fibre can the couple many cavities in a row
without a significant decrease of the light field along it.
  
\section{Acknowledgements}

The authors would like to thank Ata\c{c} Imamo\u{g}lu, Tobias Kippenberg and Kerry Vahala
for discussions and Alex Retzker for
proof-reading the manuscript. This work is part
of the QIP-IRC supported by EPSRC (GR/S82176/0), the Integrated
Project Qubit Applications (QAP) supported by the IST directorate
as Contract Number 015848' and was also supported by the Alexander
von Humboldt Foundation, the Conselho Nacional de Desenvolvimento
Cient\'ifico e Tecnol\'ogico (CNPq), Hewlett-Packard and the Royal Society.

%\bibliographystyle{unsrt}
%\bibliography{../BIB/mybib}

\end{document}